\documentstyle[multicol,prl, epsf,aps]{revtex}
\newcommand{\be}{\begin{equation}}
\newcommand{\ee}{\end{equation}}
\newcommand{\bea}{\begin{eqnarray}}
\newcommand{\eea}{\end{eqnarray}}
\newcommand{\ba}{\begin{array}}
\newcommand{\ea}{\end{array}}

\newcommand{\bb}{\bibitem}

\begin{document}
\draft
\tightenlines

\title{\bf New Universality classes for generic higher character Lifshitz points}
\author{Marcelo M. Leite\footnote{e-mail:leite@fis.ita.br}}
\address{\it Departamento de F\'\i sica, Instituto
Tecnol\'ogico de Aeron\'autica, Centro T\'ecnico Aeroespacial,
12228-900, S\~ao Jos\'e dos Campos, SP, Brazil}
%\date{March 2001}
\maketitle

%\end{center}
\vspace{0.2cm}
\begin{abstract}
{\it We describe new universality classes associated to generic higher
character Lifshitz critical behaviors for systems with arbitrary
short range competing interactions. New renormalization-group arguments 
are proposed for anisotropic and isotropic systems. The usual 
$m$-axial Lifshitz universality classes are obtained as a simple limit of 
this arbitrary competing critical behavior.}
\end{abstract}

\vspace{1cm}
%\pacs{PACS: 64.60.Kw; 64.70.-p; 64.70.Rh}

%\newpage

\begin{multicols}{2}
\narrowtext

The universality classes characterizing the critical behavior of
competing systems are natural generalizations of those describing
ordinary short-ranged critical systems. In magnetic systems \cite{Ho-Lu-Sh},
the axial next-nearest-neighbor Ising (ANNNI) model \cite{Sea}
explains this generalized critical behavior, as antiferromagnetic
exchange interactions between second neighbor spins are included along a
single axis in a cubic lattice. The uniaxial Lifshitz
point arises at the confluence of the disordered, a uniformly ordered
and a modulated phase. Second character Lifshitz points 
generalize the uniaxial one when the competing axes occur in 
arbitrary space directions. Physical applications of this behavior 
include magnetic 
modulated materials \cite{Be,We}, high-$T_{c}$ superconductors, liquid 
crystals, among others \cite{Se}. The theoretical description of this 
criticality has been put forth
using different field-theoretic tools \cite{Medi,AL,Le} and numerical
Monte Carlo simulations for the ANNNI model \cite{He}. \par
The ANNNI model can be generalized 
by including further ferromagnetic couplings between third neighbors 
along the competing axis. The phase diagram is characterized by two 
parameters varying with the temperature, namely, the ratio of exchange 
couplings between the second and the first neighbors as well as the ratio 
of exchange interactions between the third and the first neighbors. 
The projection of the temperature axis in the plane of these parameters 
results in a region of intersection where the different phases 
characterizing the system meet, known as a uniaxial Lifshitz point of 
third character \cite{Se1}. Including more alternate couplings along the 
competing direction, the number of parameters in the phase 
diagram increases. Analogously, the
uniaxial Lifshitz point of higher character is defined  at the confluence
of the distinct phases of the system. The $m_{L}$-fold Lifshitz point
of character $L$ arises when alternate couplings are permitted
up to the $L$th neighbors along $m_{L}$ directions and the ratios
of the exchange interactions between different neighbors are kept fixed 
at the associated Lifshitz temperature \cite{Se2,Ni1}. \par 
Lifshitz critical behaviors share at least two issues with a broad class 
of theories in other physical arenas. In some phenomenological models of 
quantum gravity, the Planck scale 
and the speed of light are two observer-independent scales which induce
modifications  of the usual energy-momentum dispersion relations
\cite{Ma1}. A nonlinear realization of Lorentz invariance is
maintained and can be generalized to include more than two relevant
scales \cite{Ma2}. Lorentz invariance can also be preserved in 
addressing the dark matter problem in cosmology by making use of a 
scalar field with higher derivative terms, named ``$k$-essence'' 
\cite{AMS}. Alternatively, Planck scale effects may break Lorentz 
invariance in the ultraviolet regime due to the modification of the 
dispersion relations. Effective free quantum field theories with modified 
cubic kinetic terms have just been proposed \cite{MP} to test 
this idea. Thus, a renormalization
group analysis with several independent length scales together
with new perturbative techniques to solve interacting higher derivative 
field theories are highly desirable to elucidating new effects in a 
broader context. \par
In this letter, we introduce the most general $d$-dimensional competing
system resulting in a new static critical behavior by permitting
only nearest neighbor interactions along $(d-m_{2} -...-m_{L})$ directions,
second neighbor competing interactions along $m_{2}$ directions, etc., up
to $L$th neighbor competing interactions along $m_{L}$ directions. The
new universality classes for this system generalize the higher
character universality classes in a nontrivial way. Since each
competing direction defines distinct modulated
phases, the generic higher character Lifshitz point arises at the confluence
among the disordered, a uniformly ordered and {\it several} modulated ordered
phases. The universality classes for this arbitrary competing exchange
couplings Ising model, named hereafter CECI model, depend on
$(N, d, m_{2}, ..., m_{L})$ where $N$ is the number of components of the
order parameter. The particular isotropic universalities $(d=m_{n})$ are
obtained independently. We find multiscale scaling relations using a simple 
scaling hypothesis consistent with homogeneity. Then, we develop a new 
dimensional regularization procedure to calculate Feynman diagrams and 
critical exponents for generic higher character Lifshitz points up to 
two-loop order. \par
We start with the following bare Lagrangian density:
\begin{eqnarray}
L &=& \frac{1}{2}
|\bigtriangledown_{(d- \sum_{n=2}^{L} m_{n})} \phi_0\,|^{2} +
\sum_{n=2}^{L} \frac{\sigma_{n}}{2}
|\bigtriangledown_{m_{n}}^{n} \phi_0\,|^{2} \\ \nonumber
&& + \sum_{n=2}^{L} \delta_{0n}  \frac{1}{2}
|\bigtriangledown_{m_{n}} \phi_0\,|^{2}
+ \sum_{n=3}^{L-1} \sum_{n'=2}^{n-1}\frac{1}{2} \tau_{nn'}
|\bigtriangledown_{m_{n}}^{n'} \phi_0\,|^{2} \\ \nonumber
&&+ \frac{1}{2} t_{0}\phi_0^{2} + \frac{1}{4!}\lambda_0\phi_0^{4} .
\end{eqnarray}
At the Lifshitz point, the fixed ratios among the
exchange couplings explained above translate into this field-theoretic
version though the conditions $\delta_{0n} = \tau_{n n'} =0$. All even 
momentum powers up to $2L$ become relevant in the free propagator, 
realizing a situation considered by Wilson in the 1970's \cite{Wil}. \par
The critical space dimension $d_{c}$ above which the classical (mean field)
behavior takes over the system is given by
$d_{c} = 4 + \sum_{n=2}^{L} \frac{(n-1)}{n} m_{n}$. The small parameter for
a consistent perturbative expansion is
$\epsilon_{L} = 4 + \sum_{n=2}^{L} \frac{(n-1)}{n} m_{n} - d$. \par
In momentum space, the $m_{n}$ competing axes only
possess the term $k_{(n)}^{2n}$ due to the Lifshitz condition. Thus,
we can set $\sigma_{n}=1$ and perform a dimensional redefinition of
the competing momenta defining arbitrary competing axes. Hence, if
the canonical dimension (in mass units $\Lambda$) of the quadratic
noncompeting momenta is given by $[q] = \Lambda$, then
$[k_{(n)}] = \Lambda^{\frac{1}{n}}$. Therefore, the volume
element in momentum space for arbitrary Feynman integrals has canonical
dimension $\Lambda^{d - \sum_{n=2}^{L} \frac{(n-1)}{n} m_{n}}$. Let
$\xi_{1}$, $\xi_{2}$,..., $\xi_{L}$ be the correlation lengths
characterizing arbitrary spatial directions of the system. Close to the
Lifshitz critical temperature these quantities are related among each, but
can be considered as independent relevant length scales in the problem as 
follows. \par
Consider $L$ normalization conditions for the one-particle irreducible (1PI)
vertex parts. The renormalization of the vertices at the Lifshitz 
temperature is consistent in the infrared limit if the theory is renormalized 
at nonvanishing external momenta. Let $\kappa_{n}$ be the external momenta 
scale used to fix the
renormalized theory along arbitrary space directions. For instance, for
$n=1$ the nonvanishing components of the external momenta are parallel to
the noncompeting $(d - \sum_{n=2}^{L} m_{n})$-dimensional subspace. For
$n>1$, the nonvanishing components of the external momenta occur only along
the $m_{n}$-dimensional competing subspace. Let $p_{i(n)}$ be the external
momenta associated to a generic vertex part. We can choose
several symmetry points $SP_{n}$ along arbitrary directions defined by
$p_{i(n)}.p_{j(n)}= \frac{\kappa_{n}^{2}}{4} (4 \delta_{ij} -1)$. The momentum scale of the two point function is fixed by $p_{(n)}^{2} = \kappa_{n}^{2}=1$.
The 1PI renormalized vertex parts at the Lifshitz critical temperature
are given in terms of the bare vertices as \cite{Le,amit,BLZ}$(n=1,...,L)$:
\end{multicols}
\hspace{-0.5cm}
\rule{8.7cm}{0.1mm}\rule{0.1mm}{2mm}
\widetext
\begin{eqnarray}
\Gamma_{R(n)}^{(N,L)} (p_{i (n)}, Q_{i(n)}, g_{n}, \kappa_{n})
&=& Z_{\phi (n)}^{\frac{N}{2}} Z_{\phi^{2} (n)}^{L}
(\Gamma_{(n)}^{(N,L)} (p_{i (n)}, Q_{i (n)}, \lambda_{n}, \Lambda_{n})\\ \nonumber
&& - \delta_{N,0} \delta_{L,2}
\Gamma^{(0,2)}_{(n)} (Q_{(n)}, Q_{(n)}, \lambda_{n}, \Lambda_{n})|_{Q^{2}_{(n)} = \kappa_{n}^{2}}),
\end{eqnarray}
\hspace{9.1cm}
\rule{-2mm}{0.1mm}\rule{8.7cm}{0.1mm}
\begin{multicols}{2}
\narrowtext\noindent
where $g_{n}$ are the renormalized coupling constants, $Z_{\phi (n)}$ and
$Z_{\phi^{2} (n)}$ are the field and temperature normalization constants and
$\Lambda_{n}$ are the corresponding momentum cutoffs along the noncompeting
($n=1$) and competing directions ($1< n \leq L$). Notice that $Q_{i (n)}$ are
the insertion external momenta along the different spatial directions for
vertex parts including composite $\phi^{2}(x)$ operators. \par
Our dimensional redefinitions of the momenta components characterizing
each type of competing axes implies that the renormalized and bare
dimensionful coupling constants can be expressed in terms of dimensionless
couplings as $g_{n} = u_{n}(\kappa_{n}^{2n})^{\frac{\epsilon_{L}}{2}}$ and
$\lambda_{n} = u_{0n}(\kappa_{n}^{2n})^{\frac{\epsilon_{L}}{2}}$.
Since we are going to analyze the
situation for $d< d_{c}$, we can suppress the cutoffs $\Lambda_{n}$ in
the argument of the bare vertices recalling, however, that the limit
$\Lambda_{n} \rightarrow \infty$ is implicit. The invariance of the bare
vertex functions with the renormalizaton scales $\kappa_{n}$ implies the
renormalization group (RG) equation
\end{multicols}
\hspace{-0.5cm}
\rule{8.7cm}{0.1mm}\rule{0.1mm}{2mm}
\widetext
\begin{equation}
(\kappa_{n} \frac{\partial}{\partial \kappa_{n}} +
\beta_{n}\frac{\partial}{\partial u_{n}}
- \frac{1}{2} N \gamma_{\phi (n)}(u_{n}) + L \gamma_{\phi^{2} (n)}(u_{n}))
\Gamma_{R(n)}^{(N,L)} = \delta_{N,0} \delta_{L,2} (\kappa_{n}^{-2 l})^\frac{\epsilon_{L}}{2} B_{n}(u_{n}) .
\end{equation}
\hspace{9.1cm}
\rule{-2mm}{0.1mm}\rule{8.7cm}{0.1mm}
\begin{multicols}{2}
\narrowtext\noindent
The functions $\beta_{n}, \gamma_{\phi (n)}$ and $\gamma_{\phi^{2} (n)}(u_{n})$
are calculated at fixed bare coupling $\lambda_{n}$. In terms of
dimensionless quantities, they read
\begin{mathletters}
\begin{eqnarray}
\beta_{n} &=& - n \epsilon_{L}(\frac{\partial ln u_{0n}}{\partial u_{n}})^{-1}, \\
\gamma_{\phi (n)}(u_{n})  &=&
\beta_{n} \frac{\partial ln Z_{\phi (n)}}{\partial u_{n}}, \\
\gamma_{\phi^{2} (n)}(u_{n}) &=& - \beta_{n}
\frac{\partial ln Z_{\phi^{2} (n)}}{\partial u_{n}}.
\end{eqnarray}
\end{mathletters}
At the fixed points, under flows in the external momenta
$p_{i(n)}, Q_{i(n)}$ along arbitrary space directions,
the vertex parts
$\Gamma_{R (n)}^{(N,L)} (\rho p_{i (n)}, \rho Q_{i (n)}, u_{n}^{*},
\kappa_{n})$ have a simple scaling property.
As usual, define $\eta_{n}= \gamma_{\phi(n)}^{*}$. The correlation length
exponents are given by $\nu_{n}^{-1} = 2n - \gamma_{\phi^{2} (n)}^{*} $.
The determination of the scaling relations above and below the Lifshitz
critical temperature follows closely that for the second character
Lifshitz points \cite{Le,Ref}. One finds
\begin{mathletters}
\begin{eqnarray}
&&\alpha_{n} = 2 - n(d - \sum_{j=2}^{L} \frac{(j-1)}{j} m_{j}) \nu_{n}, \\
&&\beta_{n} = \frac{1}{2} \nu_{n} (n(d-\sum_{i=2}^{L} \frac{(i-1)}{i} m_{i}) - 2n + \eta_{n}),\\
&&\gamma_{n} = \nu_{n} (2 n  - \eta_{n}),\\
&&\delta_{n} = \frac{n(d-\sum_{i=2}^{L} \frac{(i-1)}{i} m_{i}) + 2n -
\eta_{n}}{n(d-\sum_{i=2}^{L} \frac{(i-1)}{i} m_{i}) - 2n + \eta_{n}},
\end{eqnarray}
\end{mathletters}\noindent
which imply the Widom $\gamma_{n} = \beta_{n} (\delta_{n} -1)$ and
Rushbrook $\alpha_{n} + 2 \beta_{n} + \gamma_{n} = 2$ relations for
{\it arbitrary} competing or noncompeting subspaces, since $n=1,...,L$. \par
The isotropic cases are defined by $d=m_{n}$, when all competing interactions
not lying on the $m_{n}$ competing axes are switched off. The upper
critical dimension is $d_{c} =4n$ implying
that the perturbative parameter is $\epsilon_{4n} = 4n -d$. There is only
one type of correlation length $\xi_{4n}$. The canonical dimension of the
momenta in mass units is redefined to be $[k] = \Lambda^{\frac{1}{n}}$.
The volume element of Feynman integrals in momentum space has canonical
dimension $[d^{d}k] = \Lambda^{\frac{d}{n}}$. Following similar steps
as done in the anisotropic cases, we obtain the relations:
\begin{mathletters}
\begin{eqnarray}
\gamma_{4n} &=& \nu_{4n} (2n  - \eta_{4n}), \\
\alpha_{4n} &=& 2 - d \nu_{4n} \\
\delta_{4n} &=& \frac{d + 2n - \eta_{4n}}{d - 2n + \eta_{4n}}, \\
\beta_{4n} &=& \frac{1}{2} \nu_{4n} (d - 2n + \eta_{4n}),
\end{eqnarray}
\end{mathletters}\noindent
which imply the Widom $\gamma_{4n} = \beta_{4n} (\delta_{4n} -1)$ and
Rushbrook $\alpha_{4n} + 2 \beta_{4n} + \gamma_{4n} = 2$ laws. Note
that they trivially reduce to the usual $\phi^{4}$ theory for $n=1$, after
the identification $\alpha_{4} \equiv \alpha$, etc.. These
relations are the same as those found by Nicoll {\it et al.} \cite{Ni2},
with the additional scaling relation for $\beta_{n}$ obtained in the present
work in a simple manner. \par
Consider the exponents for the anisotropic cases obtained from
the critical theory defined by (2) at nonzero external momenta. At this
critical point the bare free critical propagator ($t=0$) is given by
$G_{0}^{-1} = p^{2} + \sum_{n=2}^{L} k_{n}^{2n}$, where $\vec{p}$
is a wave vector along the $(d-m_{2} -...-m_{L})$ noncompeting directions,
and $\vec{k_{n}}$ are the wave vectors along the $m_{n}$ distinct type of
competing axes. Let $k'_{i (n)}$ be the external momenta along the $m_{n}$ 
competing axes in the contribution of an arbitrary 1PI vertex part, which 
at the symmetry points discussed above can be simply
written as $k'_{(n)}$. If the loop momenta is $k_{(n)}$, we can
employ a generalized ``orthogonal approximation'' \cite{Le} to resolving the
Feynman integrals up to two-loop level with arbitrary external momenta.
Using the approximation
$(k_{(n)} + k'_{(n)})^{2n} \cong (k_{(n)}^{n} + k_{n}^{' n})^{2}$, we
can redefine the loop momenta as $K_{(n)} = k_{(n)}^{n}$. In each
integration over the competing momenta we pick out only the contribution
corresponding to a homogeneous function of the momenta.
We absorb the geometric angular factor
$[(\Pi_{n=2}^{L} \frac{1}{2n} S_{m_{n}} \Gamma(\frac{m_{n}}{2n}))
S_{(d- \sum_{n=2}^{L}m_{n})}
\Gamma(2 - \sum_{n=2}^{L}  \frac{m_{n}}{2n})]$ in a redefinition of the
coupling constant after performing each loop integral. Using (2), we
find that the fixed point is indepent of the scaling direction
considered and the critical exponents $\nu_{n}$ and $\eta_{n}$ can be
calculated diagramatically at two- and three-loop order,
respectively. The other exponents are obtained from the scaling laws
derived above. They are given by
\begin{mathletters}
\begin{eqnarray}
&& \nu_{n} = \frac{\nu_{1}}{n} = \frac{\frac{1}{2} + \frac{(N + 2)}{4(N + 8)} \epsilon_{L}
+  \frac{1}{8}\frac{(N + 2)(N^{2} + 23N + 60)} {(N + 8)^3} \epsilon_{L}^{2}}{n},\\
&& \eta_{n} = n \eta_{1}= \frac{n}{2} \epsilon_{L}^{2}\,\frac{N + 2}{(N+8)^2}
[1 + \epsilon_{L}(\frac{6(3N + 14)}{(N + 8)^{2}} - \frac{1}{4})] ,\\
&& \gamma_{n} = 1 + \frac{(N+2)}{2(N + 8)} \epsilon_{L}
+\frac{(N + 2)(N^{2} + 22N + 52)}{4(N + 8)^{3}} \epsilon_{L}^{2}, \\
&& \alpha_{n} = \frac{(4 - N)}{2(N + 8)} \epsilon_{L}
- \frac{(N + 2)(N^{2} + 30N + 56)}{4(N + 8)^{3}} \epsilon_{L}^{2} ,\\
&& \beta_{n} = \frac{1}{2} - \frac{3}{2(N + 8)} \epsilon_{L}
+ \frac{(N + 2)(2N + 1)}{2(N + 8)^{3}} \epsilon_{L}^{2} ,\\
&& \delta_{n} = 3 + \epsilon_{L}
+ \frac{(N^{2} + 14N + 60)}{2(N + 8)^{2}} \epsilon_{L}^{2}.
\end{eqnarray}
\end{mathletters}
Notice that the anisotropic exponents
$\gamma_{n}, \alpha_{n}, \beta_{n}$ and $ \delta_{n}$ easily reproduce
those of the second character Lifshitz behavior when
$m_{3} = ... = m_{L} = 0$ with $m_{2} = m \neq 0$,
which turn into the Ising-like exponents for  $m_{2} = m = 0$. The 
generalized universality classes exposed here and the properties 
$\nu_{n} = \frac{\nu_{1}}{n}$ and $\eta_{n}=
n \eta_{1}$, when $m_{2} = ...= m_{L-1}=0$ with $m_{L} \neq 0$, imply 
that our anisotropic scaling relations are equivalent to those found in
Ref. \cite{Ni1}, except for the scaling relation of the magnetization
exponent which was not obtained there. Strong 
anisotropic scaling \cite{he} is {\it exact} at the loop order considered 
here in the most general anisotropic situation, and might be expected to 
hold at arbitrary loop levels. \par
Using the same approximation scheme to perform Feynman integrals as
outlined above, the isotropic exponents are even simpler to obtain.
The angular factor to be absorbed in the loop integrals is simply
the area of the $m_{n}$-dimensional sphere $S_{m_{n}}$. The critical
exponents are given by \cite{ref}:
\end{multicols}
\hspace{-0.5cm}
\rule{8.7cm}{0.1mm}\rule{0.1mm}{2mm}
\widetext
\begin{mathletters}
\begin{eqnarray}
&& \eta_{4n} = \frac{1}{2n} \epsilon_{4n}^{2}\,\frac{N + 2}{(N+8)^2}
[1 + \frac{\epsilon_{4n}}{n}(\frac{6(3N + 14)}{(N + 8)^{2}} - \frac{1}{4})] ,\\
&&\nu_{4n} = {\frac{1}{2n} + \frac{(N + 2)}{4n^{2}(N + 8)} \epsilon_{4n}
+  \frac{1}{8n^{3}}\frac{(N + 2)(N^{2} + 23N + 60)} {(N + 8)^3} \epsilon_{4n}^{2}},\\
&& \gamma_{4n} = 1 + \frac{(N+2)}{2n(N + 8)} \epsilon_{4n}
+\frac{(N + 2)(N^{2} + 22N + 52)}{4n^{2} (N + 8)^{3}} \epsilon_{4n}^{2}, \\
&& \alpha_{4n} = \frac{(4 - N)}{2n(N + 8)} \epsilon_{4n}
- \frac{(N + 2)(N^{2} + 30N + 56)}{4n^{2}(N + 8)^{3}} \epsilon_{4n}^{2} ,\\
&& \beta_{4n} = \frac{1}{2} - \frac{3}{2n(N + 8)} \epsilon_{4n}
+ \frac{(N + 2)(2N + 1)}{2n^{2}(N + 8)^{3}} \epsilon_{4n}^{2} ,\\
&& \delta_{4n} = 3 + \frac{\epsilon_{4n}}{n}
+ \frac{(N^{2} + 14N + 60)}{2n^{2}(N + 8)^{2}} \epsilon_{4n}^{2}.
\end{eqnarray}
\end{mathletters}
\hspace{9.1cm}
\rule{-2mm}{0.1mm}\rule{8.7cm}{0.1mm}
\begin{multicols}{2}
\narrowtext\noindent
The isotropic exponents $\alpha_{4n}, \beta_{4n}, \gamma_{4n}$ and
$\delta_{4n}$ depend {\it explicitly} on the number of
neighbors coupled through competing exchange interactions
$n$ and labels the isotropic universality classes. These critical
exponents reduce to those from the Ising-like ($\phi^{4}$)
universality class for $n=1$. For $n \geq 2$, these exponents can never
be obtained from those of the anisotropic cases described above, since
the latter do not depend explicitly on $n$. This confirms that the
isotropic universality classes are independent from the anisotropic
universality classes for Lifshitz points of generic higher character. \par
In summary, the CECI model proposed here presents the most general sort 
of second order phase transition for arbitrary types of competition and 
realizes new universality classes for generic higher character Lifshitz 
critical behaviors. The scaling relations are obtained through a generalized 
scaling hypothesis which defines independent renormalization group
tranformations in each competing (or noncompeting) subspace. The critical 
exponents are calculated up to two-loop level using a new analytical 
dimensional regularization method to compute Feynman integrals with 
arbitrary even momentum powers in the propagator, which may be fruitful 
in quantum field theory applications. Universal quantities as 
amplitude ratios \cite{le1,le2} can be pursued in 
this framework. We relegate the details of the results described here for 
a subsequent publication \cite{le3}.\par
It is important to mention that a 
new anisotropic behavior emerges when $d = m_{2}+...+m_{L}$. The noncompeting 
directions which tend to favor the existence of the ferromagnetic phase, are 
absent in this new situation. This particular behavior deserves a complete and 
careful study for future investigations. The analysis of the most general 
crossover effects as well as the tricritical behavior of
the model are also worthwhile. \par
I would like to thank S. K. Adhikari and S. R. Salinas for useful
comments on the manuscript and financial support from FAPESP, grant 
number 00/06572-6.

\end{multicols}

\end{document}